\documentclass{ifacconf}

\usepackage{graphicx}
\usepackage{natbib} 
\usepackage[dvipsnames]{xcolor}
\usepackage{amsfonts,amsmath,amssymb}
\usepackage[utf8]{inputenc}
\usepackage{epstopdf}
\usepackage{soul}
\usepackage{url}
\usepackage{todonotes}
\usepackage{float}
\usepackage[font=small]{caption, subcaption}
\usepackage[draft,inline,nomargin,index]{fixme}

\definecolor{forest}{HTML}{009933}
\definecolor{bluesea}{HTML}{6666ff}
\definecolor{rederror}{HTML}{ff3333}
\fxsetup{theme=color,mode=multiuser}
\FXRegisterAuthor{lg}{lgenv}{\color{forest}LG} 
\FXRegisterAuthor{mb}{mbenv}{\color{bluesea}MB} 
\FXRegisterAuthor{lm}{lmenv}{\color{rederror}LM} 

\newcommand{\ts}{\zeta}           
\newcommand{\im}{\eta}           
\newcommand{\ee}{e}              
\newcommand{\zz}{z}              
\newcommand{\yy}{y}              
\newcommand{\oo}{\hat{\ee}}      
\newcommand{\ext}{\xi}           
\newcommand{\oxt}{\hat{\ext}}    
\newcommand{\is}{\varsigma}      
\newcommand{\io}{\theta}         
\newcommand{\ks}{\kappa_s}       
\newcommand{\iss}{\Sigma}        
\newcommand{\ios}{\Theta}        
\newcommand{\pr}{\psi}           
\newcommand{\idf}{\varphi}       
\newcommand{\ido}{\vartheta}     
\newcommand{\hg}{l}              
\newcommand{\dg}{\Lambda(\hg)}   
\newcommand{\hh}{H}              
\newcommand{\ff}{\gamma}         
\newcommand{\vdp}{\chi}         
\newcommand{\qq}{q}
\newcommand{\bb}{b}
\newcommand{\bs}{\boldsymbol{\bb}}
\newcommand{\xx}{\boldsymbol{x}}
\newcommand{\by}{\boldsymbol{y}}
\newcommand{\RR}{\mathbb{R}}
\newcommand{\NN}{\mathbb{N}}
\newcommand{\MM}{\mathcal{M}}
\newcommand{\CC}{\mathcal{C}}

\newcommand{\DD}{\mathcal{D}}
\newcommand{\EE}{\mathcal{E}}
\newcommand{\DS}{\mathcal{DS}}
\newcommand{\HH}{\mathcal{H}}
\newcommand{\JJ}{\mathcal{J}}
\newcommand{\KL}{\mathcal{KL}}
\newcommand{\KK}{\mathcal{K}}

\newcommand{\Aa}{\mathcal{A}}
\newcommand{\Bb}{\mathcal{B}}
\newcommand{\TT}{\text{T}}
\newcommand{\loss}{\mathcal{J}}
\newcommand{\wset}{\mathcal{W}}
\newcommand{\piset}{\mathcal{P}}
\newcommand{\zd}{f_0}
\newcommand{\kf}{\kappa}
\newcommand{\kfs}{\mathcal{X}}
\newcommand{\km}{\boldsymbol{\mathcal{K}}}
\newcommand{\kv}{\boldsymbol{\kappa}}
\newcommand{\ys}{\boldsymbol{u}}
\newcommand{\xs}{\boldsymbol{\eta}}
\newcommand{\pmu}{\mu}           
\newcommand{\prior}{m}           
\newcommand{\pvar}{\sigma^2}     
\newcommand{\nn}{\sigma_n^2}     
\newcommand{\np}{\sigma_p^2}
\newcommand{\thr}{\sigma_{\text{thr}}^2}
\newcommand{\T}{^{\top}}
\newcommand{\s}{^{\star}}
\newcommand{\lp}{\left(}
\newcommand{\rp}{\right)}

\newcommand\figref{Figure~\ref}
\newcommand\figsref{Figures~\ref}
\newcommand\secref{Section~\ref}

\newcommand\eqqref{Equation~\eqref}
\newcommand\lref{Lemma~\ref}
\newcommand\assmref{Assumption~\ref}
\newcommand\assmsref{Assumptions~\ref}
\newcommand\clref{Claim~\ref}

\begin{document}
\begin{frontmatter}
\title{Data-driven Output Regulation via Gaussian Processes and Luenberger Internal Models}
\author[First]{Lorenzo Gentilini}
\author[Second]{Michelangelo Bin}
\author[First]{Lorenzo Marconi}
\address[First]{Dept.~of Electrical, Electronic and Information Engineering, \\
Alma Mater Studiorum University of Bologna, Bologna, Italy \\
(e-mails: \{lorenzo.gentilini6, lorenzo.marconi\}@unibo.it).}
\address[Second]{Dept.~of Electrical and Electronic Engineering, \\
Imperial College London, London, UK \\
(e-mail: m.bin@imperial.ac.uk).}

\begin{abstract}
   This paper deals with the problem of adaptive output regulation for multivariable nonlinear systems by presenting a
   learning-based adaptive internal model-based design strategy.
   The approach builds on the recently proposed adaptive internal model design techniques based on the theory of
   nonlinear Luenberger observers, and the adaptation side is approached as a probabilistic regression problem.
   In particular, Gaussian processes priors are employed to cope with the learning problem.
   Unlike the previous approaches in the field, here only coarse assumptions about the friend structure are required, making the
   proposed approach suitable for applications where the exosystem is highly uncertain.
   The paper presents performance bounds on the attained regulation error and numerical simulations showing how
   the proposed method outperforms previous approaches.
\end{abstract}
\end{frontmatter}

\section{Introduction}%
\label{sec:INTRODUCTION}
\vspace*{-0.3cm}
In this paper we consider a class of nonlinear systems of the form
\begin{equation}%
   \label{eq:GENERAL-SYSTEM}
   \begin{matrix}
      \dot{x} = f \lp x,w,u \rp, & \ee = h \lp x,w \rp,
   \end{matrix}
\end{equation}
with state $x \in \RR^{n_{x}}$, control input $u \in \RR^{n_{u}}$, regulation error $\ee \in \RR^{n_{\ee}}$, and with
$w \in \RR^{n_{w}}$ an exogenous signal.
As customary in the literature of output regulation, we assume that the exogenous signal $w$ belongs to the set of
solutions of an exosystem of the form
\begin{equation}%
   \label{eq:EXOGENOUS-SYSTEM}
   \dot{w} = s(w),
\end{equation}
originating in a compact invariant subset $\wset$ of $\RR^{n_{w}}$.
For the class of systems~\eqref{eq:GENERAL-SYSTEM}~\eqref{eq:EXOGENOUS-SYSTEM}, in this paper we consider the
problem of design an output-feedback regulator of the form
\begin{equation*}
   \begin{matrix}
      \dot{x}_c = f_x \lp x_c, \ee \rp, & u_c = k_c \lp x_c, \ee \rp,
   \end{matrix}
\end{equation*}
that ensures boundedness of the closed-loop trajectories and asymptotically removes the effect of $w$
from the regulated output $\ee$, thus ideally obtaining $\ee \lp t \rp \rightarrow 0$ as $t \rightarrow \infty$.
More precisely, the sought regulator must ensure
\begin{equation*}
   \limsup_{t \rightarrow \infty} \left\| \ee \lp t \rp \right\| \le \epsilon 
\end{equation*}
with $\epsilon \ge 0$ possibly a small number measuring the regulator's asymptotic performance.
In this work we focus on the specific case of \emph{adaptive approximate regulation}, where the 
aforementioned control objective is relaxed to the case with $\epsilon > 0$, and a learning technique
is employed to cope with uncertainties in the exosystem, and in the plant dynamics.
In particular, the adaptation side is approached in a system identification fashion where
the Gaussian process regression is used to infer the internal model dynamics directly out of the collected data.
\vspace*{-0.5cm}
\subsubsection*{Related Works}
Most of the work in the field of output regulation can be traced-back to~\cite{francis1976internal} and~\cite{davison1976robust} who firstly formalize and solve
the asymptotic regulation problem in the context of linear systems.
Asymptotic results was given also in the field of Single-Input-Single-Output (SISO) nonlinear systems,
first in a local context (\cite{byrnes1997structurally},~\cite{isidori1990output}),
and later in a purely nonlinear framework (\cite{byrnes2003limit},~\cite{marconi2007output}),
based on the ``non-equilibrium'' theory (\cite{byrnes2003asymptotic}).
Recently, asymptotic regulators have been also extended to some classes of multivariable nonlinear
systems (\cite{wang2016nonlinear},~\cite{wang2017robust}).
The major drawbacks of asymptotic regulators reside in their complexity and fragility (\cite{bin2018robustness}).
Indeed, sufficient conditions under which asymptotic regulation is ensured are typically expressed by equations
whose analytic solution becomes a hard task even for relative simple problems. Moreover, even if a regulator can
be build, asymptotic regulation may be lost at front of exosystem perturbation and plant uncertainties.
The aforementioned problems motivates the researchers to move toward more robust solutions, introducing the concept of adaptive and approximate regulation.
Among the approaches to approximate regulation it is worth mentioning~\cite{marconi2008uniform} and~\cite{astolfi2015approximate}, whereas practical regulators
can be found in~\cite{isidori2012robust} and~\cite{freidovich2008performance}. Adaptive designs of regulators can be found
in~\cite{priscoli2006new} and~\cite{pyrkin2017output}, where linearly parametrized
internal models are constructed in the context of adaptive control, in~\cite{bin2019adaptive} where discrete-time adaptation algorithms
are used in the context of multivariable linear systems, and in~\cite{forte2016robust},~\cite{bin2019class},~\cite{bin2020approximate}, where adaptation of a nonlinear
internal model is approached as a system identification problem.

Learning dynamics models is also an active research topic.
In particular, Gaussian Processes (GPs) are increasingly used to estimate unknown dynamics (\cite{kocijan2016modelling},~\cite{buisson2020actively}).
Unlike other nonparametric models, GPs represent an attractive tool in learning dynamics due to their flexibility in modeling nonlinearities and
the possibility to incorporate prior knowledge (\cite{rasmussen2003gaussian}).
Moreover, since GPs allow for analytical formulations, theoretical guarantees on the a posteriori can be drawn directly from the collected
data (\cite{umlauft2020learning},~\cite{lederer2019uniform}).
Recently, GP models spread inside the field of nonlinear optimal control (\cite{sforni2021learning}), with several applications to the particular case of
Model Predictive Control (MPC) (\cite{torrente2021data},~\cite{kabzan2019learning}), and inside the field of nonlinear observers (\cite{buisson2021joint}).

\subsubsection*{Contributions}
In this paper, we propose a data-driven adaptive output regulation scheme, built on top of the recently published
works~\cite{bin2020approximate} and~\cite{gentilini2022adaptive}, in which the problem of approximate regulation is solved
by means of a regulator embedding an adaptive internal model.
Unlike previous approaches, here the high flexibility of Gaussian process priors (\cite{rasmussen2003gaussian}) is used to adapt
an internal model unit in a discrete-time system identification fashion, enabling
the possibility to handle a possibly infinite class of input signals needed to ensure zero regulation error
(the so-called \emph{friend},~\cite{isidori1990output}).
Compared to~\cite{bin2020approximate}, where the identifier is related to a particular choice of class of functions,
to which the friend may (or may not) belongs, the proposed approach aims to perform
probabilistic inference in a possibly infinite-dimensional space.
Unlike~\cite{gentilini2022adaptive}, the proposed regulator relies on non-high-gain stabilising actions and
Luenberger-like internal models that lead to a fixed choice of the model order.
The latter property, jointly with the black-box nature of Gaussian process methods, makes the proposed approach suitable for
those applications where the exosystem dynamics is highly uncertain and the friend structure is not a priori known.
Theoretical performance bounds on the attained regulation error are analytically established.

The paper unfolds as follows. In~\secref{sec:PROBLEM-SET-UP} we briefly describe the problem at hand along with the standing assumptions over the presented results build.
\secref{sec:APPROXIMATE-REGULATION} reviews the most recent advancements in the output regulation field, and introduces the barebone regulator adapted for this work,
while~\secref{sec:GAUSSIAN-INFERENCE} introduces the basics of Gaussian process inference. In~\secref{sec:PROPOSED-SOLUTION} we present the proposed regulator and state
the main result of the paper. Finally, in~\secref{sec:NUMERICAL-SIMULATION} a numerical example is presented.

\section{Problem set-up \& Preliminaries}%
\label{sec:PROBLEM-SET-UP}
\vspace*{-0.3cm}
In this section, we first detail the subclass of problems that this work focuses on, along with the constructive assumptions.
Then, a Luenberger-like internal model design technique is reviewed, together with the adaptive regulator of~\cite{bin2020approximate}.
Finally, basic concepts behind the notion of Gaussian process regression are introduced.

\subsection{Approximate Nonlinear Regulation}%
\label{sec:APPROXIMATE-REGULATION}
\vspace*{-0.3cm}
In this paper, we focus on a subclass of the general regulation problem presented in~\secref{sec:INTRODUCTION}, by considering systems of the form
\begin{equation}%
   \label{eq:NORMAL-FORM}
   \begin{split}
      &\dot{\zz} = \zd \lp w, \zz, \ee \rp, \\
      &\dot{\ee} = A\ee + B \lp \qq \lp w, \zz, \ee \rp + \bb \lp w, \zz, \ee \rp u \rp, \\
      &\yy = C\ee,
   \end{split}
\end{equation}
in which $\zz \in \RR^{n_{\zz}}$ together with the error dynamics $\ee \in \RR^{n_{\ee}}$ represent the overall state of the plant.
The quantities $u \in \RR^{n_{\yy}}$ and $\yy \in \RR^{n_{\yy}}$ are the control input and the measured output respectively, while $w \in \RR^{n_{w}}$ is
an exogenous input, $\zd : \RR^{n_{w}} \times \RR^{n_{\zz}} \times \RR^{n_{\ee}} \mapsto \RR^{n_{\zz}}$,
$\qq : \RR^{n_{w}} \times \RR^{n_{\zz}} \times \RR^{n_{\ee}} \mapsto \RR^{n_{\yy}}$, $\bb : \RR^{n_{w}} \times \RR^{n_{\zz}} \times \RR^{n_{\ee}} \mapsto \RR^{n_{\yy} \times n_{\yy}}$
are continuous functions, and $A$, $B$, and $C$ are defined as
\begin{equation*}
   \begin{matrix}
      A =
      \begin{pmatrix}
         0_{(r-1)n_{\yy} \times n_{\yy}} & I_{(r-1)n_{\yy}} \\
         0_{n_{\yy} \times n_{\yy}} & 0_{n_{\yy} \times (r-1)n_{\yy}}
      \end{pmatrix}, &
      B =
      \begin{pmatrix}
         0_{(r-1)n_{\yy} \times n_{\yy}} \\
         I_{n_{\yy}}
      \end{pmatrix},
   \end{matrix}
\end{equation*}
\begin{equation*}
   C =
   \begin{pmatrix}
      I_{n_{\yy}} & 0_{n_{\yy} \times (r-1)n_{\yy}}
   \end{pmatrix},
\end{equation*}
for some $r \in \NN$, consisting in a chain of $r$ integrators of dimension $n_{\yy}$.
\setcounter{thm}{0}
The aforementioned framework embraces a large number of use-cases addressed in literature.
In particular, all systems presenting a well-defined vector relative degree and admitting a canonical normal form,
or that are strongly invertible and feedback linearisable fit inside the proposed framework.
Nevertheless, this approach limits to systems having an equal number of inputs and controlled outputs $\lp n_{\yy} \rp$.
The results presented in the next sections are grounded over the following set of standing assumptions.
\setcounter{thm}{0}
\begin{assum}%
   \label{assum:CONTINUITY-ASSUMPTION}
   The function $\zd$ is locally Lipschitz and the functions $\qq$ and $\bb$ are $\CC^1$ functions, with local Lipschitz derivative.
\end{assum}
\begin{assum}%
   \label{assum:MINIMUM-PHASE-ASSUMPTION}
   There exists a $\CC^1$ map $\pi: \piset \subset \RR^{n_{w}} \mapsto \RR^{n_{z}}$, with $\piset$ an open neighborhood of $\wset$, satisfying
   \begin{equation*}
      L_{s \lp w \rp}^{\lp w \rp} \pi \lp w \rp = \zd \lp w, \pi \lp w \rp, 0 \rp,
   \end{equation*} 
   with $L_{s \lp w \rp}^{\lp w \rp} \pi \lp w \rp = \frac{\partial \pi (w)}{\partial w} s(w)$, such that the system
   \begin{equation*}
      \begin{matrix}
         \dot{w} = s \lp w \rp, & \dot{\zz} = \zd \lp w, \zz, \ee \rp,
      \end{matrix}
   \end{equation*}
   is Input-to-State Stable (ISS) with respect to the input $\ee$, relative to the compact set $\Aa = \left\{ \lp w, \zz \rp \in \wset \times \RR^{n_{\zz}} : \zz = \pi \lp w \rp \right\}$.
\end{assum}
\begin{assum}%
   \label{assum:STABILIZABILITY-ASSUMPTION}
   There exists a known constant nonsingular matrix $\bs \in \RR^{n_{\yy} \times n_{\yy}}$ such that the inequality
   \begin{equation*}
      \left\| (\bb(w, \zz, \ee) - \bs)\bs^{-1} \right\| \le 1 - \mu_0,
   \end{equation*}
   holds for some known scalar $\mu_0 \in \lp 0, 1 \rp$, and for all $\lp w, \zz, \ee \rp \in \wset \times \RR^{n_{\zz}} \times \RR^{n_{\ee}}$.
\end{assum}
\setcounter{thm}{1}
\begin{rem}
   Although not necessary (see~\cite{byrnes2003limit}),~\assmref{assum:MINIMUM-PHASE-ASSUMPTION} is a minimum-phase assumption 
   customary made in the literature of output regulation (see~\cite{isidori2017lectures},~\cite{pavlov2006uniform}).
   In particular,~\assmref{assum:MINIMUM-PHASE-ASSUMPTION} is asking that the zero dynamics
   \begin{equation*}
      \begin{matrix}
         \dot{w} = s(w), & \dot{\zz} = \zd \lp w, \zz, 0 \rp,
      \end{matrix}
   \end{equation*}
   has a steady-state of the kind $\zz = \pi(w)$, compatible with the control objective $y = 0$.
   As a consequence, the ideal input $u\s$ making the set $\Bb = \Aa \times \{0\}$ invariant for~\eqref{eq:NORMAL-FORM} reads as
   \begin{equation*}
      u\s \lp w, \pi(w) \rp = - \bb \lp w, \pi(w), 0 \rp^{-1} \qq \lp w, \pi(w), 0 \rp.
   \end{equation*}
   The ability of the regulator to generate such an input is generally referred to as the \emph{internal model property}.
   With a little abuse of notation, from now on we refer to $ u\s \lp w, \pi(w) \rp$ with $ u\s \lp w \rp$.
\end{rem}
\begin{rem}
   \assmref{assum:CONTINUITY-ASSUMPTION} asks for some Lipschitz conditions on maps that play a
   fundamental role in the stability analysis. In particular, Lipschitz continuity is required as long as high-gain-based observers are
   employed inside the regulator structure, later detailed in~\eqref{eq:PROPOSED-REGULATOR}.
   Furthermore, even if in this work we deal with data-driven adaptive control techniques, that ideally require smoothness assumptions
   on the function to be identified $u\s$, in practice the adaptation of the internal model structure proposed by~\cite{marconi2007output}
   makes the problem solvable without any further assumption. The details about this issue are more deeply discussed in~\secref{sec:PROPOSED-SOLUTION}.
\end{rem}
\begin{rem}
   \assmref{assum:STABILIZABILITY-ASSUMPTION} is a stabilizability assumption asking that $\bb(w, \zz, \ee)$ is always invertible whatever $(x, \zz, \ee)$ is (see~\cite{wang2017robust}).
   Moreover, the designer is required to have access to an estimate $\bs$ of $\bb(w, \zz, \ee)$ which captures enough information about its behavior.
\end{rem}
In this framework, we now recall two results based on~\cite{marconi2007output} and~\cite[Theorem 1]{bin2020approximate}.
\setcounter{thm}{0}
\begin{lem}%
   \label{lem:LUMBERGER-INTERNAL-MODEL}
   Let~\assmref{assum:MINIMUM-PHASE-ASSUMPTION} holds and let $n_{\im} = 2(n_{w}+n_{\zz}+1)$.
   Then, for any choice of controllable pair $\lp F, G \rp$, with $F$ a Hurwitz matrix, there exist
   two maps $\tau: \RR^{n_{w}} \mapsto \RR^{n_{\im}}$, and $\ff: \RR^{n_{\im}} \mapsto \RR^{n_{\yy}}$ such that
   for all $w$ in $\wset$
   \begin{equation*}
      \begin{split}
      \ff \circ \tau(w) &= u\s(w), \\
      L_{s(w)}^{(w)} \tau(w) &= F \tau(w) + G u\s(w),
      \end{split}
   \end{equation*}
   and the system
   \begin{equation*}
      \begin{split}
         \dot{w} & = s(w), \\
         \dot{\zz} & = \zd(w,\zz,\ee), \\
         \dot{\im} & = F \im + G u\s(w) + \delta,
      \end{split}
   \end{equation*}
   is ISS relative to the set $\EE = \big\{ \lp w, \zz, \im \rp \in \Aa \times \RR^{n_{\im}}: \im = \tau(w) \big\}$ and with respect to the input $(\ee, \delta)$.

\end{lem}

Let~\assmref{assum:CONTINUITY-ASSUMPTION},~\ref{assum:MINIMUM-PHASE-ASSUMPTION}, and~\ref{assum:STABILIZABILITY-ASSUMPTION} hold,
and let $\MM = \left\{ \pr(\io, \cdot): \RR^{n_{\im}} \mapsto \RR^{n_{\yy}} | \io \in \ios \right\}$, with $\ios$ a finite-dimensional normed vector space,
be a finite-dimensional model set where $\ff$ is supposed to range. Consider the following regulator structure\footnote{Same regulator proposed
in~\cite{bin2020approximate}, with the only difference in the definition of $\io = \ido \lp \is \rp$. It is, in fact, equivalent to set
$\io^{+} = \ido \lp \is \rp$ with $\dot{\io} = 0$ and delay the optimality condition of one step, i.e. $\io\s \in \arg\min_{\io \in \ios} \loss \lp j-1, \io \rp$.}
\begin{equation*}
   \begin{split}
      &
      \begin{cases}
         \dot{\ts} = 1 \\
         \dot{\im} = F \im + G u \\
         \dot{\oo} = A \oo + B \lp \oxt + \bs u \rp + \dg \hh \lp \yy - \oo_1 \rp \\
         \dot{\oxt} = - \bs \pr \lp \io, \im, u \rp + \hg^{r+1} \hh_{r+1} \lp \yy - \oo_1 \rp\\
         \dot{\is} = 0
      \end{cases} \\
      & \lp \ts, \im, \oo, \oxt, \is, \io, \yy \rp \in C_{\ts} \times \RR^{n_{\im} + n_{\ee} + n_{\yy}} \times \iss \times \ios \times \RR^{n_{\yy}},
   \end{split}
\end{equation*}
\begin{equation*}
   \begin{split}
      &
      \begin{cases}
         \ts^{+} = 0 \\
         \im^{+} = \im \\
         \oo^{+} = \oo \\
         \oxt^{+} = \oxt\\
         \is^{+} = \idf \lp \is, \im, u \rp
      \end{cases} \\
      & \lp \ts, \im, \oo, \oxt, \is, \io, \yy \rp \in D_{\ts} \times \RR^{n_{\im} + n_{\ee} + n_{\yy}} \times \iss \times \ios \times \RR^{n_{\yy}},
   \end{split}
\end{equation*}
with $\io = \ido \lp \is \rp$ and output $u = \bs^{-1}\text{sat}(-\oxt + \ks \lp \oo \rp)$.
Where $A$, $B$, $\bs$ are the same in~\eqref{eq:NORMAL-FORM} and~\assmref{assum:STABILIZABILITY-ASSUMPTION},
while $(F,G)$ and $n_{\im}$ are the same of~\lref{lem:LUMBERGER-INTERNAL-MODEL}, and
$\iss$ finite-dimensional normed vector space. The sets $C_{\ts}$, $D_{\ts}$ are defined as
\begin{equation*}
   \begin{matrix}
      C_{\ts} = \left[ 0, \overline{\TT} \right], &
      D_{\ts} = \left[ \underline{\TT}, \overline{\TT} \right],
   \end{matrix}
\end{equation*}
with $\overline{T}$, $\underline{T} \in \RR_{+}$, satisfying $0 < \underline{\TT} \le \overline{\TT}$.
Furthermore, $\dg = \text{diag} \lp \hg I_{n_{\yy}}, \hg^{2} I_{n_{\yy}}, \dots, \hg^{r} I_{n_{\yy}} \rp$, $\hh = \text{diag} \lp \hh_1, \dots, \hh_r \rp$, and
$\hh_i = \text{diag} \lp h^1_i, \dots, h^{n_{\yy}}_i \rp$ with $\{h_{1}^j, h_{2}^j, \dots, h_{r+1}^j\}$ for all
$j = 1, \dots, n_{\yy}$ coefficients of a Hurwitz polinomial, and $\hg \in \RR_{>0}$ is a control parameter.
Let the tuple $\lp \MM, \iss, \pr, \ios, \ido \rp$ be such that the \emph{identifier requirements}, relative to a given cost function $\loss$, are satisfied.
Namely there exist $\beta_{\is} \in \KL$, locally Lipschitz $\rho_{\is}, \rho_{\io} \in \KK$, a compact set $\iss\s \subset \iss$ and,
for each solution pair $((\ts, w, \is, \io),(d_{\im}, d_{\yy}))$ to
\begin{equation}%
   \label{eq:PERTURBED}
   \begin{split}
      &
      \begin{cases}
         \dot{\ts} = 1 \\
         \dot{w} = s(w) \\
         \dot{\is} = 0
      \end{cases} \\
      & \lp \ts, w, \is, \io, d_{\im}, d_{\yy} \rp \in C_{\ts} \times W \times \iss \times \ios \times \RR^{n_{\im}} \times \RR^{n_{\yy}}, \\
      &
      \begin{cases}
         \ts^{+} = 0 \\
         w^{+} = w \\
         \is^{+} = \idf \lp \is, \tau(w) + d_{\im}, \ff(\tau(w)) + d_{\yy} \rp
      \end{cases} \\
      & \lp \ts, w, \is, \io, d_{\im}, d_{\yy} \rp \in D_{\ts} \times W \times \iss \times \ios \times \RR^{n_{\im}} \times \RR^{n_{\yy}}, \\
   \end{split}
\end{equation}
with $\io = \ido \lp \is \rp$, there exists a pair $(\is\s, \io\s)$ and a $j\s \in \NN$, such that $((\ts, w, \is\s, \io\s),(0,0))$ is a solution pair to~\eqref{eq:PERTURBED}
satisfying $\is\s(j) \in \iss\s$ for all $j \ge j\s$ and the following properties hold:
\begin{enumerate}
   \item \emph{Optimality:}
   For each $j \ge j\s$
   \begin{equation*}
      \io\s \in \arg\min_{\io \in \ios} \loss \lp j, \io \rp.
   \end{equation*}
   \item \emph{Stability:}
   For each $j$
   \begin{equation*}
      \begin{split}
         \left| \is(j) - \is\s(j) \right| \le \max & \Big\{ \beta_{\is}\lp \is(0) - \is\s(0), j \rp, \\
         & \hspace{1.2cm} \rho_{\is}\lp \left| \lp d_{\im}, d_{\yy} \rp \right| \rp \Big\}.
      \end{split}
   \end{equation*}
   \item \emph{Regularity:}
   The function $\ido$ satisfies 
   \begin{equation*}
      \left| \ido\lp \is \rp - \ido\lp \is\s \rp \right| \le \rho_{\io} \lp \left| \is - \is\s \right| \rp,
   \end{equation*}
   for all $\lp \is, \is\s \rp \in \iss \times \iss\s$, the map $\pr \lp \io, \im, u \rp$ is $\CC^1$ with locally Lipschitz derivative in the argument $\im$.
\end{enumerate}
Then, for each compact sets $Z_0 \subset \RR^{n_{\zz}}$, $E_0 \subset \RR^{n_{\ee}}$, and $S_0 \subset \RR^{n_{\ee}} \times \RR^{n_{n_{\yy}}}$
of initial conditions for $\zz$, $\ee$, and $\lp \oo, \bs \rp$ respectively, there exists $\hg_{s}\s > 0$ such that if $\hg > \hg_{s}\s$
then the aggregate state $\xx = (\ts, w, \zz, \ee, \im, \oo, \oxt, \is, \io)$ of the closed-loop system is bounded.
Moreover, there exists a $\alpha_{\xx} > 0$ and for each $\underline{\TT} > 0$, an $\hg_{\epsilon}\s > 0$, such that if
$\hg > \hg_{\epsilon}\s \lp \underline{\TT} \rp$ then
\begin{equation*}
   \limsup_{t \rightarrow \infty} \left| y(t) \right| \le \alpha_{\xx} \limsup_{t+j \rightarrow \infty} \left| u\s(w) - \pr(\io\s, \tau(w)) \right|.
\end{equation*}

\setcounter{thm}{4}
\begin{rem}
   We stress the fact that building an identifier satisfying the requirements presented in~\secref{sec:APPROXIMATE-REGULATION}
   is necessary linked to a specific choice of the model set $\MM$, which is the
   space of functions where $\ff$ is supposed to range. Due to implementation constraints, it is customary to focus on finite-dimensional sets,
   which allows the parametrization of $\ff$ by a parameter $\io$ ranging in a finite-dimensional vector space $\ios$.
   This, in turn, limits the flexibility of the proposed approach, especially when the structure of the friend is not a priori known.
   For this reason, we drop the assumption about $\MM$ by performing regression in the space of \emph{universal approximators} made by
   Gaussian processes.
\end{rem}
\subsection{Gaussian Process Inference}%
\label{sec:GAUSSIAN-INFERENCE}
\vspace*{-0.3cm}
The key idea behind the proposed approach consists in modeling the unknown function $\ff$ as the realization of a Gaussian process.
A GP is a stochastic process such that any finite number of outputs is assigned a joint Gaussian distribution with prior mean function
$\prior: \RR^{n_{\im}} \mapsto \RR^{n_{\yy}}$ and covariance defined through the kernel $\kf: \RR^{n_{\im}} \times \RR^{n_{\im}} \mapsto \RR$
(\cite{rasmussen2003gaussian}).
While there are many possible choices of mean and covariance functions, in this work we keep the formulation of $\kf$ general, with
the only constraint expressed by~\assmref{assm:K-CONTINUOUS-BOUNDED} below. 
Yet we force, without loss of generality, $\prior \lp \im \rp = 0_{n_{\yy}}$ for any $\im \in \RR^{n_{\im}}$.
Thus, we assume that
\begin{equation*}
   \ff \sim \mathcal{GP} \lp 0, \kf \lp \cdot, \cdot \rp \rp.
\end{equation*}
Supposing to have access to a data-set of samples collected at different time instants $t_i \in \RR_{>0}$,
$\DS = \{\lp \im, u \rp \in \RR^{n_{\im}} \times \RR^{n_{\yy}} : \im = \im(t_i), u = u(t_i) \text{ with } i = 1,\dots,N\}$,
with each pair $\lp\im, u\rp \in \DS$ obtained as $u(t_i) = \ff(\im(t_i)) + \varepsilon(t_i)$ with
$\varepsilon(t_i) \sim \mathcal{N}(0, \nn I_{n_{\yy}})$ white Gaussian noise with known variance $\nn$, the regression is performed by conditioning
the prior GP distribution on the training data $\DS$ and a test point $\im$.
Denoting $\xs = (\im(t_1), \dots, \im(t_N))\T$ and $\ys = (u(t_1), \dots, u(t_N))\T$, the conditional posterior distribution given the data-set is still a
Gaussian process with mean $\pmu$ and variance $\pvar$ given by (\cite{rasmussen2003gaussian})
\begin{equation}%
   \label{eq:GP-POSTERIOR}
   \begin{split}
      \pmu \lp \im \rp & = \kv \lp \im \rp\T \lp \km + \nn I_{N} \rp^{-1} \ys, \\
      \pvar \lp \im \rp & = \kf(\im, \im) - \kv(\eta)\T\left(\km + \nn I_{N}\right)^{-1}\kv(\eta),
   \end{split}
\end{equation}
where $\km \in \RR^{N \times N}$ is the \textit{Gram matrix} whose $(k,h)$-th entry is $\km_{k,h} = \kf(\xs_k, \xs_h)$,
with $\xs_k$ the $k$-th entry of $\xs$, and $\kv(\im) \in \RR^N$ is the kernel vector whose $k$-th component is $\kv_k(\im) = \kf(\im, \xs_k)$.
The problem of inferring an unknown function from a finite set of noisy data can be seen as a special case of \emph{ridge regression} where the
prior assumptions (mean and covariance) are encoded in terms of \emph{smoothness} of $\pmu$.
In particular, let $\HH$ be a RKHS associated with the kernel function $\kf$, then an estimation of $\ff$ can be inferred by minimizing the functional
\begin{equation}%
   \label{eq:COST-FUNCTIONAL}
   \JJ = \frac{\lambda_s}{2} \left\| \pmu \right\|^2_{\HH} + Q(\ys, \pmu(\xs)),
\end{equation}
where $\left\| \pmu \right\|^2_{\HH}$ is the RKHS norm and represents the smoothness assumptions on $\pmu$ 
(this term plays the role of \emph{regularizer}), while $Q$ assesses the quality of the prediction
$\pmu(\xs)$ with respect to the observed data $\ys$ (\cite{rasmussen2003gaussian}).
According to the \textit{Representer Theorem} (\cite{o1986automatic}), each minimizer $\pmu \in \HH$ of $\JJ$ takes the form  
$\pmu(\im) = \kv(\im)\alpha$, with $\alpha$ which depends on the particular choice of the prediction error.
In the particular case in which $Q(\ys, \pmu(\xs))$ corresponds to a negative log-likelihood
of a Gaussian model with variance $\nn$, namely
\begin{equation*}
   Q(\ys, \pmu(\xs)) = \frac{1}{2 \nn} \left\| \ys - \pmu(\xs) \right\|^2_2,
\end{equation*}
the value of $\alpha$ recovers the expression in Equation~\eqref{eq:GP-POSTERIOR} as
\begin{equation*}
   \alpha = (\km + \nn I_{N})^{-1}\ys.
\end{equation*}
From now on we suppose that the following standing assumptions hold (see~\cite{buisson2021joint},~\cite{lederer2021uniform})
\setcounter{thm}{3}
\begin{assum}%
   \label{assm:GAMMA-CONTINUOUS-BOUNDED}
   The unknown function $\ff$ has a bounded norm in the RKHS $\HH$ generated to the kernel $\kf$.
\end{assum}
\begin{assum}%
   \label{assm:K-CONTINUOUS-BOUNDED}
   The kernel function $\kf$ is \emph{isotropic}\footnote{Isotropic kernels are functions depending only on the Euclidean distance of their arguments.
   In this respect, the compact notation $\kf \lp x, x' \rp = \kf\lp \left\| x-x' \right\| \rp$ is commonly used.}
   and Lipschitz continuous with constant $L_{\kf}$, with a locally Lipschitz derivative of constant $L_{d\kf}$.
\end{assum}
Although any kernel fulfilling~\assmref{assm:K-CONTINUOUS-BOUNDED} can be a valid candidate, in the following,
we exploit the commonly adopted \textit{squared exponential kernel} as prior covariance function, which can be expressed as
\begin{equation}%
   \label{eq:EXPONENTIAL-KERNEL}
   \kf(\im, \im') = \np \exp \lp -\lp \im - \im' \rp \T \Lambda^{-1}  \lp \im - \im'\rp \rp
\end{equation}
for all $\im, \im' \in \RR^{n_{\im}}$, where $\Lambda = \text{diag}(2\lambda_{\im_1}^2, \dots, 2\lambda_{\im_{n_{\im}}}^2)$, $\lambda_{\im_{i}} \in \RR_{>0}$ is
known as \emph{characteristic length scale} relative to the $i$-th signal, and $\np$ is usually called
\emph{amplitude} (\cite{rasmussen2003gaussian}).
\setcounter{thm}{5}
\begin{rem}
   \assmref{assm:K-CONTINUOUS-BOUNDED} is asking some Lipschitz continuity property of the unknown function that makes it well-representable by means of a Gaussian process prior.
   Nevertheless, it represents a very strong assumption, difficult to be checked even if the unknown function is known.
   \assmref{assm:K-CONTINUOUS-BOUNDED} can be relaxed to the condition that $\ff$ is a sample from the Gaussian process $\mathcal{GP} \lp 0, \kf \lp \cdot, \cdot \rp \rp$,
   which, in turn, leads to a larger pool of posssible unkown functions and it is easier to be check. As an example, the pool generated by
   the squared exponenial kernel~\eqqref{eq:EXPONENTIAL-KERNEL} is equal to the space of continuous functions.
\end{rem}
\begin{rem}
   The isotropic kernel structure is a customary (although not necessary) assumption in the literature of Gaussian process regression.
   In this respect, the following results can be generalized for any Lipschitz continuous kernel by means of well-known arguments (see~\cite{lederer2021uniform}).
\end{rem}
We conclude this section by recalling two results based on~\cite{lederer2021uniform}.
\setcounter{thm}{2}
\begin{lem}%
   \label{lem:VARIANCE-BOUND}
   Consider a zero-mean Gaussian process defined through a kernel $\kf: \kfs \times \kfs \mapsto \RR$, satisfying~\assmref{assm:K-CONTINUOUS-BOUNDED}
   on a compact subset $\kfs$ of $\RR^{n_{\im}}$, and $N \in \NN$ observations
   $\DS = \left\{ \lp x^1, y^1 \rp, \dots, \lp x^N, y^N \rp \right\}$, with
   $y^i = f \lp x^i \rp + \varepsilon^i$, where $\varepsilon^i \sim \mathcal{N}(0, \nn I_{n_{\yy}})$.
   Then, the posterior variance is bounded as
   \begin{equation*}
      \pvar \lp x \rp \le \kf(0) - \frac{\kf(\rho)^2}{\kf(0) + \frac{\nn}{\left| \Bb_{\rho}(x) \right|}} \hspace*{0.5cm} \forall x \in \kfs,
   \end{equation*}
   where $\Bb_{\rho}(x) = \left\{ x' \in \DS : \left\| x - x' \right\| \le \rho \right\}$
   denotes the training data-set restricted to a ball around $x$ with radius $\rho \in \RR_{>0}$, and $\left| \cdot \right|$ denotes the cardinality.
\end{lem}
\begin{lem}%
   \label{lem:MEAN-BOUND}
   Consider a zero-mean Gaussian process defined through a kernel $\kf: \kfs \times \kfs \mapsto \RR$, satisfying~\assmref{assm:K-CONTINUOUS-BOUNDED}
   on the compact set $\kfs$. Furthermore, consider a continuous unknown function $f: \kfs \mapsto \RR$ with Lipschitz constant $L_f$, and $N \in \NN$ observations
   $y^i = f \lp x^i \rp + \varepsilon^i$, with $\varepsilon^i \sim \mathcal{N}(0, \nn I_{n_{\yy}})$.
   Then, there exists $\rho \in \RR_{>0}$ such that the posterior mean $\pmu$ and posterior variance $\pvar$ conditioned on the training data
   $\DS = \left\{ \lp x^1, y^1 \rp, \dots, \lp x^N, y^N \rp \right\}$
   are continuous with Lipschitz constants $L_{\pmu}$ and $L_{\pvar}$ on $\kfs$, respectively, satisfying
   \begin{equation*}
      \begin{split}
         & L_{\pmu} \le L_{\kf} \sqrt{N} \left\| \lp \km + \nn I_N \rp^{-1} \by \right\|, \\
         & L_{\pvar} \le 2 \rho L_{\kf} \lp 1 + N \left\| \lp \km + \nn I_N \rp^{-1} \right\| \max_{x, x' \in \kfs} \kf(x,x') \rp,
      \end{split}
   \end{equation*}
   with $\xx = (x^1, \dots, x^N)\T$ and $\by = (y^1, \dots, y^N)\T$.
   Moreover, pick $\delta \in \lp 0,1 \rp$ and set
   \begin{equation*}
      \begin{split}
         & \beta \lp \rho \rp = 2 \log \lp \frac{M \lp \rho, \kfs \rp}{\delta} \rp, \\
         & \alpha \lp \rho \rp = \lp L_{f} + L_{\pmu} \rp \rho + \sqrt{\beta \lp \rho \rp L_{\pvar} \rho},
      \end{split}
   \end{equation*}
   with $M \lp \rho, \kfs \rp$ the $\rho$-covering number~\footnote{The minimum number satisfying
   $\min_{x \in \kfs} \max_{x' \in \DS} \left\| x - x' \right\| \le \rho$.} related to the set $\kfs$.
   Then, the bound
   \begin{equation*}
      \left| f(x) - \pmu(x) \right| \le \sqrt{\beta \lp \rho \rp} \pvar \lp x \rp + \alpha \lp \rho \rp \hspace*{0.5cm} \forall x \in \kfs
   \end{equation*}
   holds with probability al least $1 - \delta$.
\end{lem}

\section{The Proposed Regulator}%
\label{sec:PROPOSED-SOLUTION}
\vspace*{-0.3cm}
The proposed regulator reads as follows
\begin{equation*}%
   \label{eq:PROPOSED-REGULATOR}
   \begin{split}
      &
      \begin{cases}
         \dot{\ts} = 1 \\
         \dot{\im} = F \im + G u \\
         \dot{\oo} = A \oo + B \lp \oxt + \bs u \rp + \dg \hh \lp \yy - \oo_1 \rp \\
         \dot{\oxt} = - \bs \lp \dot{\pmu} \lp \im, \is, \io \rp \rp + \hg^{r+1} \hh_{r+1} \lp \yy - \oo_1 \rp \\
         \dot{\is} = 0
      \end{cases} \\
      & \lp \ts, \im, \oo, \oxt, \is, \io, \yy \rp \in \CC, \\
      &
      \begin{cases}
         \ts^{+} = 0 \\
         \im^{+} = \im \\
         \oo^{+} = \oo \\
         \oxt^{+} = \oxt \\
         \is^{+} = \lp S \otimes I_N \rp \is +  \lp B \otimes I_N \rp \begin{bmatrix} \im & u \end{bmatrix}\T
      \end{cases} \\
      & \lp \ts, \im, \oo, \oxt, \is, \io, \yy \rp \in \DD,
   \end{split} 
\end{equation*}
with $\io = \ido \lp \is \rp$ and output $u = \bs^{-1}\text{sat}(-\oxt + \ks \lp \oo \rp)$.
Where $A$, $B$, and $\bs$ are the same as in~\eqref{eq:NORMAL-FORM} and~\assmref{assum:STABILIZABILITY-ASSUMPTION}, $F$, $G$, and $n_{\im}$ are
the same as~\lref{lem:LUMBERGER-INTERNAL-MODEL}, and $\dg$, $\hh$ are defined as in~\secref{sec:APPROXIMATE-REGULATION} with $\hg \in \RR_{>0}$ a
free control parameter fixed later to a sufficiently large number, while the matrices
$S \in \RR^{N(n_{\im} + n_{\yy}) \times N(n_{\im} + n_{\yy})}$ and $B \in \RR^{N(n_{\im} + n_{\yy})}$
have the shift form, denoting $n_{\is} = N(n_{\im} + n_{\yy})$
\begin{equation*}
   \begin{matrix}
      S = 
      \begin{pmatrix}
         0_{(n_{\is}-1) \times 1} & I_{n_{\is}-1} \\
         0 & 0_{1 \times (n_{\is}-1)}
      \end{pmatrix}, &
      B = 
      \begin{pmatrix}
         0_{(n_{\is}-1) \times 1} \\ 1
      \end{pmatrix}.
   \end{matrix}
\end{equation*}
The flow and jump set are defined as
$\CC = \big\{(\ts, \im, \oo, \oxt, \is, \io, \yy) \in \RR_{>0} \times \RR^{n_{\im} + n_{\ee} + n_{\yy}} \times \iss \times \ios \times \RR^{n_{\yy}} : 0 \le \ts \le \overline{\TT}, \pvar \lp \im, \is, \io \rp \le \thr \big\}$
and
$\DD = \big\{(\ts, \im, \oo, \oxt, \is, \io, \yy) \in \RR_{>0} \times \RR^{n_{\im} + n_{\ee} + n_{\yy}} \times \iss \times \ios \times \RR^{n_{\yy}} : \underline{\TT} \le \ts \le \overline{\TT}, \pvar \lp \im, \is, \io \rp \ge \thr \big\}$
respectively, where $\iss \subset \RR^{N(n_{\im}+n_{\yy})}$, $\ios \subset \RR^{N}$ with $N \in \NN_{>0}$, and $\underline{\TT}, \overline{\TT}, \thr \in \RR_{>0}$ satisfying $\underline{\TT} \le \overline{\TT}$ and $\np\nn(\np + \nn)^{-1} < \thr \le \np$.
The functions $\pmu \lp \im, \is, \io \rp$ and $\pvar \lp \im, \is, \io \rp$ are the a posteriori GP estimate mean and variance, respectively, after the collection of
$N$ samples. According to~\secref{sec:GAUSSIAN-INFERENCE}, denoting $\is = (\is_{\im}, \is_{u})\T$, the latter functions read to
\begin{equation*}
   \begin{split}
      & \pmu \lp \im, \is, \io \rp = \kv \lp \im \rp\T \io, \\
      & \pvar \lp \im, \is, \io \rp = \kf \lp \im, \im \rp - \kv \lp \im \rp\T \lp \km + \nn I_{N} \rp^{-1} \kv \lp \im \rp
   \end{split}
\end{equation*}
with $\io = \lp \km + \nn I_{N} \rp^{-1} \is_{u}$. In this settings, $\km$ and $\kv$ are evaluated with respect to the data-set $\DS$ as defined in~\secref{sec:GAUSSIAN-INFERENCE}.
\setcounter{thm}{0}
\begin{claim}
   Let~\assmsref{assm:GAMMA-CONTINUOUS-BOUNDED} and~\ref{assm:K-CONTINUOUS-BOUNDED} hold, then the tuple $\lp \iss, \pmu, \ido \rp$
   satisfies the identifier requirements relative to the functional~\eqref{eq:COST-FUNCTIONAL}.
\end{claim}
\begin{claim}%
   \label{claim:REGULATION-BOUND}
   Let~\assmsref{assum:CONTINUITY-ASSUMPTION}-\ref{assm:K-CONTINUOUS-BOUNDED} hold and consider the regulator~\eqref{eq:PROPOSED-REGULATOR}, then
   for each compact sets $Z_0$, $E_0$, and $S_0$
   there exists $\alpha_{\xx} > 0$ and $\hg_{\epsilon}\s > 0$, and for any choice of $\underline{\TT}, \thr \in\ \RR_{>0}$, and $N \in \NN_{>0}$,
   and for each initial condition $w_0 \in \wset$, a $\rho\s \lp w_0 \rp > 0$ such that 
   if $\hg > \hg_{\epsilon}\s$, then the bound
   \begin{equation*}
      \limsup_{t \rightarrow \infty} \left| y(t) \right| \le \alpha_{\xx} \left| \sqrt{\beta} \left[ \kf(0) - \frac{\kf(\rho\s)^2}{\kf(0) + \nn} \right] + \alpha \lp \rho\s \rp \right|
   \end{equation*}
   with $\beta$ and $\alpha$ defined as
   \begin{equation*}
      \begin{matrix}
         \beta = 2 \log \lp \frac{N}{\delta} \rp, &
         \alpha \lp \rho\s \rp = \lp L_{f} + L_{\pmu} \rp \rho\s + \sqrt{\beta L_{\pvar} \rho\s},
      \end{matrix}
   \end{equation*}
   holds with probability at least $1-\delta$.
\end{claim}
\setcounter{thm}{7}
\begin{rem}
   The quantity $\rho\s$ in~\clref{claim:REGULATION-BOUND} represents a notion of coverage of the set $\EE$ by the collected data-set.
   In particular, the lower $\rho\s$ is, the better the set $\EE$ is covered.
   As long as it approaches to zero, the regulation error approaches the lower bound
   \begin{equation*}
      \limsup_{t \rightarrow \infty} \left| y(t) \right| \le \alpha_{\xx} \left| \sqrt{\beta} \left[ \frac{\nn}{\kf(0) + \nn} \right] \right|,
   \end{equation*}
   driven by the measurement noise $\nn$.
\end{rem}

\section{Numerical Simulation}%
\begin{figure}[t]
	\centering
	\includegraphics[width=0.35\textwidth]{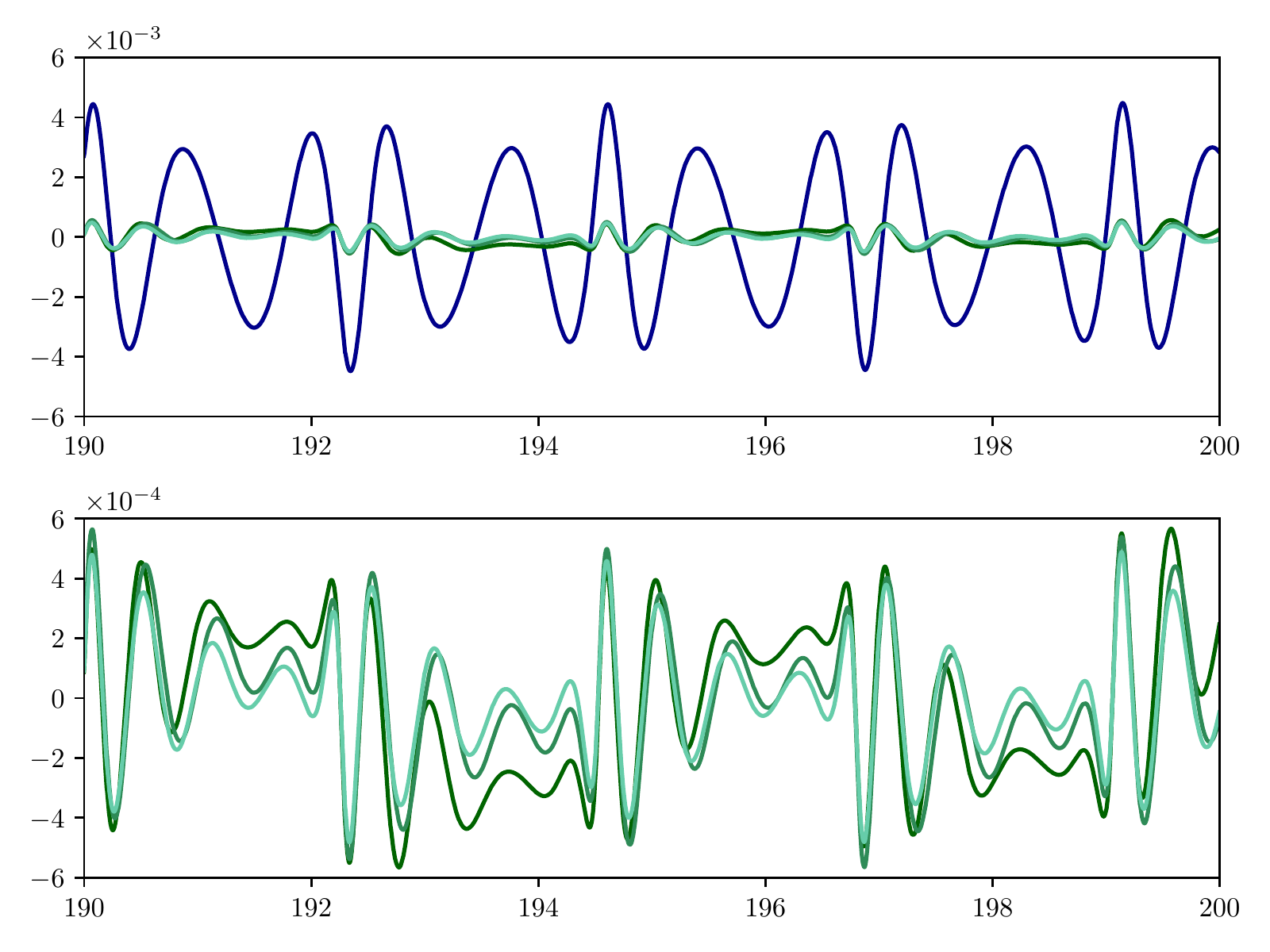}
	\caption{\textbf{Top:} steady-state evolution of the tracking error $y(t)$ in the two cases obtained by employing a linear identifier (blue line),
            and a gaussian-based identifier with $N = 50$ (dark green line), $N = 100$ (green line), and $N = 200$ (light green line).
            \textbf{Bottom:} zoom-in to highlight the error behavior.
            The reported quantities are plotted with respect to the time in seconds (abscissa).}%
   \label{fig:ZOOM-ERRORS}
\end{figure}

\begin{figure}[t]
	\centering
	\includegraphics[width=0.35\textwidth]{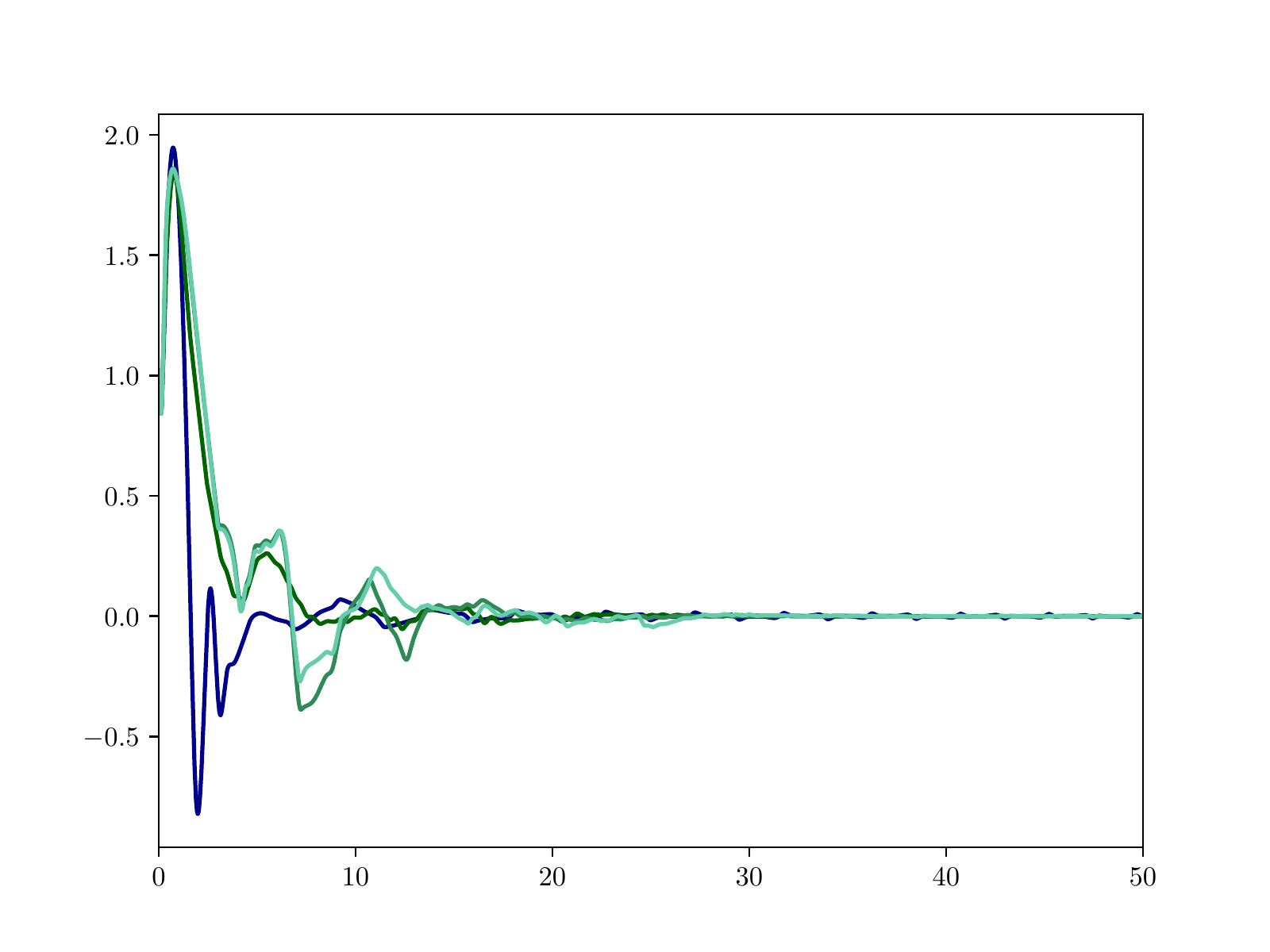}
	\caption{The transient evolution of the tracking error $y(t)$ in the two cases obtained by employing a linear identifier (blue line),
   and a gaussian-based identifier with $N = 50$ (dark green line), $N = 100$ (green line), and $N = 200$ (light green line).
   The reported quantities are plotted with respect to the time in seconds (abscissa).}%
   \label{fig:ERRORS}
\end{figure}

\begin{figure}[t]
	\centering
	\includegraphics[width=0.35\textwidth]{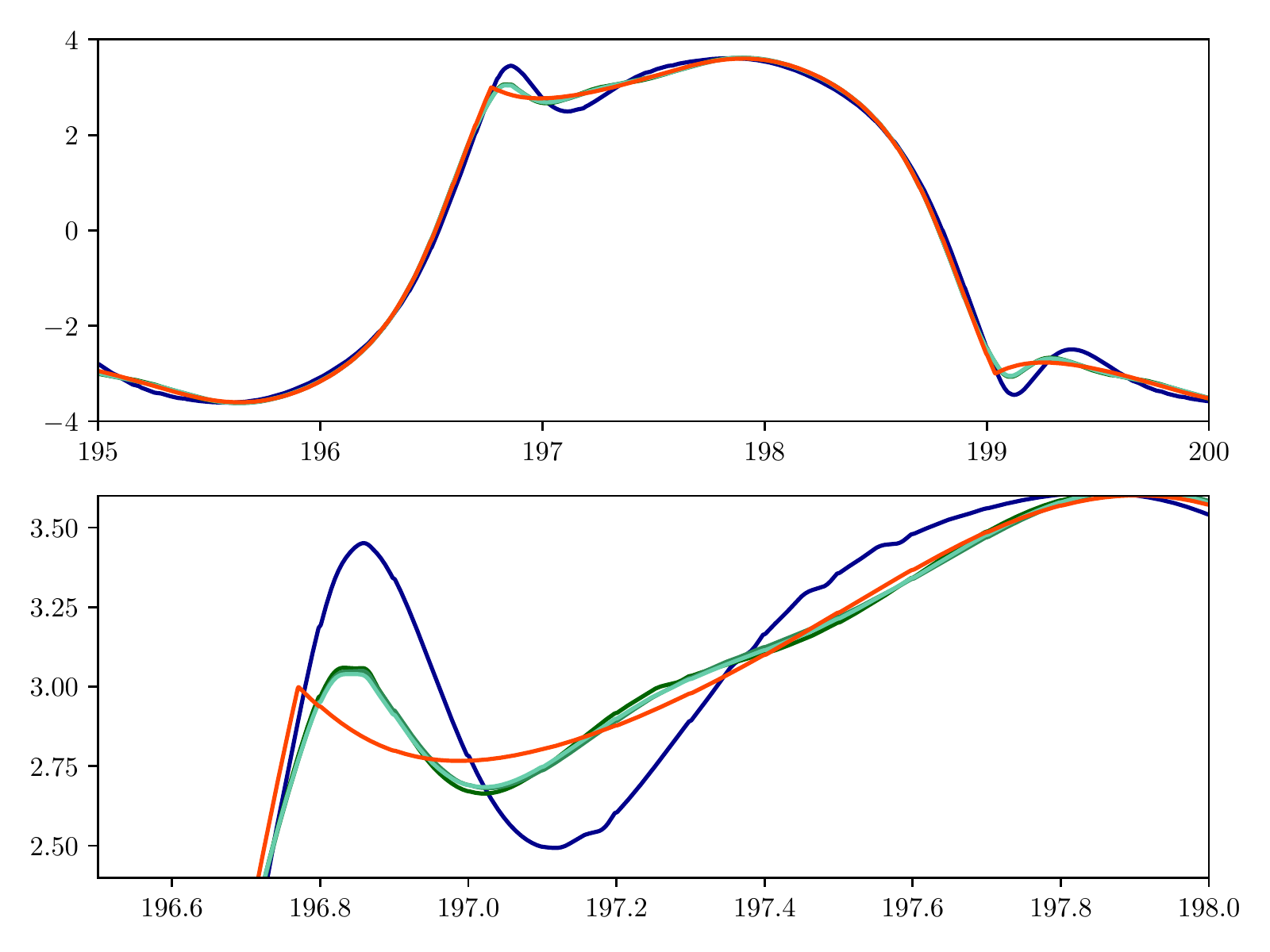}
	\caption{\textbf{Top:} Value of the real feedforward term $u\s\lp w(t) \rp$ (orange line) and of its approximations $\hat{\ff} \lp \eta(t) \rp$,
            along the system trajectory. The blue line shows the steady-state friend provied by the linear identifier,
            while the green lines report the gaussian-based identifier estimation with $N = 50$ (dark green line),
            $N = 100$ (green line), and $N = 200$ (light green line).
            \textbf{Bottom:} zoom-in to highlight the difference in the case of gaussian-based identifier with different number of samples.
            The reported quantities are plotted with respect to the time in seconds (abscissa).}%
   \label{fig:ZOOM-CTRL-ACTIONS}
\end{figure}

\label{sec:NUMERICAL-SIMULATION}
\vspace*{-0.3cm}
To test the proposed regulator performances against state-of-the-art output regulation solutions, we consider the same problem
proposed by~\cite{bin2020approximate} where the output of a Van der Pol oscillator, with unknown parameter, must be synchronized with a 
triangular wave with unknown frequency.
The forced Van der Pol oscillator is described by the following equations
\begin{equation}%
   \label{eq:VAN-DER-POL}
   \begin{split}
      & \dot{\vdp}_1 = \vdp_2, \\
      & \dot{\vdp}_2 = - \vdp_1 + a \lp 1-\vdp_1^2 \rp \vdp_2 + u,
   \end{split}
\end{equation}
with $a$ scalar unknown parameter regulating the system damping.
Furthermore, a triangular wave can be generated by an exosystem of the form
\begin{equation*}
   \begin{matrix}
      \dot{w}_1 = w_2, & \dot{w}_2 = - \varrho w_1,
   \end{matrix}
\end{equation*}
with output
\begin{equation*}
   \vdp\s(w) = \frac{2 \sqrt{w_1^2 + w_2^2}}{\pi} \arcsin \lp \frac{w_1}{\sqrt{w_1^2 + w_2^2}} \rp,
\end{equation*}
with scalar parameter $\varrho$ the unknown oscillating frequency.
The goal is to steer the output $\vdp_1$ of~\eqref{eq:VAN-DER-POL} to the reference $\vdp\s(w)$.
The error coordinates $\ee$ are thus defined as
\begin{equation*}
   \begin{pmatrix}
      \ee_1 \\ \ee_2
   \end{pmatrix} =
   \begin{pmatrix}
      \vdp_1 - \vdp\s(w) \\ \vdp_2 - L_{s(w)} \vdp\s(w)
   \end{pmatrix},
\end{equation*}
and the error system reads as
\begin{equation}%
   \label{eq:ERROR-SYSTEM}
   \begin{split}
      & \dot{\ee}_1 = \ee_2, \\
      & \dot{\ee}_2 = -\ee_1 - \vdp\s - L_{s(w)}^2 \vdp\s \\
      & \hspace{1.0cm} + a \lp 1 - \lp \ee_1 + \vdp\s(w)\rp^2 \rp \lp \ee_2 + L_{s(w)}\vdp\s(w) \rp.
   \end{split}
\end{equation}
The system~\eqref{eq:ERROR-SYSTEM} is in the same form of~\eqref{eq:NORMAL-FORM} with~\assmref{assum:MINIMUM-PHASE-ASSUMPTION} trivially fulfilled
since the $z$ dynamics is absent. Furthermore,~\assmref{assum:CONTINUITY-ASSUMPTION} and~\assmref{assum:STABILIZABILITY-ASSUMPTION} hold with
$\bs = 1$ and any $\mu \in (0,1)$. To be compliant with the results presented by~\cite{bin2020approximate} we exploit the same controller parameters
\begin{enumerate}
   \item $\ks\lp \oo \rp = K\oo$ with $K$ such that $\sigma \lp A - BK \rp = \{-1, -2\}$, and the input $u$ has been saturated inside the interval $[-100, 100]$.
   \item The internal model dimension is $n_{\im} = 2(n_{w}+1) = 6$, and the matrices $F$ and $G$ has been fixed as
   \begin{equation*}
      \begin{matrix}
         F =
         \begin{pmatrix}
            -1 & 1 & 0 & 0 & 0 & 0 \\
            0 & -1 & 1 & 0 & 0 & 0 \\
            0 & 0 & -1 & 1 & 0 & 0 \\
            0 & 0 & 0 & -1 & 1 & 0 \\
            0 & 0 & 0 & 0 & -1 & 1 \\
            0 & 0 & 0 & 0 & 0 & -1
         \end{pmatrix}, &
         G = 
         \begin{pmatrix}
            0 \\ 0 \\ 0 \\ 0 \\ 0 \\ 1
         \end{pmatrix}.
      \end{matrix}
   \end{equation*}
   \item The control parameters has been chosen as $\hg = 20$, $h_1 = 6$, $h_2 = 11$, $h_3 = 6$, and $\underline{\TT} = \overline{\TT} = 0.1$.
\end{enumerate}
The simulations reported in~\figsref{fig:ZOOM-ERRORS},~\ref{fig:ERRORS}, and~\ref{fig:ZOOM-CTRL-ACTIONS} show the proposed regulator applied
with $a = \varrho = 2$ in three cases with $N = 50$, $N = 100$, and $N = 200$. The obtained results are then compared with the regulator proposed
by~\cite{bin2020approximate} where the identifier is chosen as a least-squares identifier working on the model set
$\MM = \left\{ \pr(\io, \im): \RR^{n_{\im}} \mapsto \RR^{n_{\yy}} | \pr(\io, \im) = \io\T\im, \io \in \ios \subset \RR^{n_{\im}} \right\}$.
In all simulations the GP parameters has been kept fixed at $\nn = 0.01$ and $\thr = \np = 1$, while the kernel hyperparameters
$\boldsymbol{\lambda} = (\lambda_{\im_{1}}, \dots, \lambda_{\im_{n_{\im}}})$ has been estimated via log-likelihood minimization (\cite{rasmussen2003gaussian})
yielding to the values of $\boldsymbol{\lambda} = \lp 7.7, 34.3, 19.9, 0.4, 133.6, 1.2 \rp$. As emerges from~\figref{fig:ZOOM-ERRORS}, the proposed approach
reduces the maximum error of more than $100$ times compared to the case with least-square identifier.

\section{Conclusion}%
\label{sec:CONCLUSIONS}
\vspace*{-0.3cm}
We presented a learning-based technique to design internal model-based regulators for a large class of nonlinear systems.
The flexibility of the proposed approach makes the regulator able to deal with highly uncertain shapes of the optimal steady-state control input
useful to make zero the output error. Thanks to the fact that only coarse and qualitative knowledge about the friend is required, the proposed approach
may be employed as solution to many of the output regulation problems addressed in literature. 
The paper also derives probabilistic bounds on the attained performances and presents numerical simulations showing
how the proposed method outperforms previous approaches when the regulated plant or the exogenous disturbances are subject to unmodeled perturbations.
Future research directions will be aimed at exploring deeper the Gaussian process flexibility by focusing on
the injection of possibly a priori knowledge of the friend structure, and at investigating how the proposed performance
bound changes. We also aim to investigate if Gaussian process-based internal models may deal with non minimum-phase systems.


\bibliography{root}
\end{document}